\documentclass[cits,a4paper]{PoS}
\title{Large-$N$ mesons}
\ShortTitle{Large-$N$ mesons}
\author{\speaker{Gunnar S.\ Bali}\thanks{Adjunct Faculty:
Tata Institute of Fundamental Research, Homi Bhabha Road, Mumbai 400005, India.}, Luca Castagnini\\
        Institute f\"ur Theoretische Physik,  University of Regensburg,
93040 Regensburg, Germany\\
        E-mail: \email{gunnar.bali@ur.de}, \email{luca.castagnini@ur.de}}
\author{Biagio Lucini\\
College of Science, Swansea University, Singleton Park, Swansea SA2 8PP, UK\\
E-mail: \email{b.lucini@swansea.ac.uk}}
\author{Marco Panero\\
Instituto de F\'isica Te\'orica UAM/CSIC,
Universidad Aut\'onoma de Madrid,
Ciudad Universitaria de Cantoblanco, 28049 Madrid,
Spain\\
E-mail: \email{marco.panero@inv.uam.es}}
\abstract{We present an update of our project of
computing the meson spectrum and decay
constants in large-$N$ QCD. The results are obtained in the quenched
approximation with the Wilson fermion action
for $N=2,3,4,5,6,7$ and $17$ and extrapolated to $N=\infty$.
We non-perturbatively determine the renormalization factors
for local quark bilinears that are needed
to compute the decay constants. We
extrapolate our SU(7) results
to the continuum limit, employing four different
lattice spacings.
\vspace{4cm}

\begin{flushright}
IFT-UAM/CSIC-13-129
\end{flushright}}

\FullConference{31st International Symposium on Lattice Field Theory - LATTICE 2013\\
		July 29 - August 3, 2013\\
		Mainz, Germany}
\begin{document}
\section{Introduction}
Quantum Chromodynamics (QCD), the theory of strong interactions,
is characterized by local $\mathrm{SU}(N)$
gauge invariance where $N=3$ denotes the number of ``colours''.
The adjoint gauge bosons (gluons) couple $n_f$ ``flavours'' of
fermionic matter fields in the fundamental representation
(quarks). QCD dynamically generates a mass gap. Moreover,
at low temperatures, the (approximate) chiral symmetry is
broken. These and other non-perturbative low energy features
can be addressed systematically by lattice QCD simulations.

A different non-perturbative approach to QCD is based on an expansion
in powers of $1/N$ of the inverse number of colours~\cite{largeN}.
In the 't~Hooft limit where $N$ is sent to infinity,
keeping the 't~Hooft coupling
$\lambda=N g^2$ ($g$ denotes the gauge
coupling) as well as $n_f$ fixed,
the theory 
simplifies considerably, see ref.~\cite{reviews} for a recent review.
For instance, all amplitudes of physical processes are determined
by a particular subset of Feynman diagrams (planar diagrams),
the low-energy spectrum consists of stable meson and glueball states
and the scattering matrix becomes trivial.
One may study the physical $N=3$ case, expanding around the large-$N$
limit in terms of $1/N$. Interestingly,
the non-flavour-singlet spectra of QCD with sea quarks and quenched QCD
agree within 10~\%~\cite{Fodor:2012gf}. This may indicate that both $n_f/N$ and
$1/N^2$ corrections are small in these channels.

Another non-perturbative approach to low-energy
properties of non-Abelian gauge theories is based on the
conjectured correspondence between
gauge theories and classical gravity in
an anti-de-Sitter spacetime
(AdS/CFT correspondence)~\cite{gaugestringduality}. 
Unlike lattice regularization, in this case the continuous
spacetime symmetry is retained but the large-$N$ limit
(as well as a large 't~Hooft coupling) is implied. During the
last decade techniques based on this
correspondence have been employed
to construct models which reproduce the main features
of the meson spectrum of QCD, see, e.g., ref.~\cite{Erdmenger:2007cm}.

The large-$N$ limit also plays a central role in the chiral effective
theory approach where the $N$-dependence of
low-energy constants is known~\cite{Sharpe:1992ft} and,
within this framework, in studies of properties of unstable resonances,
see, e.g., refs.~\cite{Harada:1995dc,Pelaez:2006nj,Nieves:2011gb}.
Clearly, it is important to determine
the meson spectrum of large-$N$ QCD to constrain effective
field theory parameters and also to enable a comparison with AdS/CFT and
AdS/QCD predictions.

Large-$N$ QCD still remains far from trivial and requires lattice
simulation. The quenched theory becomes unitary and identical to
full QCD in the large-$N$ limit where quark loop effects are suppressed.
Neglecting the fermion determinant does not only
save computer time but the quenched theory should converge more rapidly
(with leading $1/N^2$ rather than with $n_f/N$ corrections)
towards the limit $N\rightarrow\infty$.
Recently, the dependence of various quantities on
$N$ was studied in quenched lattice simulations. For instance, 
pseudoscalar and vector meson masses (among other observables)
were determined in
refs.~\cite{DelDebbio:2007wk, Bali:2008an, Hietanen:2009tu, DeGrand:2012hd, Bali:2013fpo}.

In ref.~\cite{Bali:2013fpo} we chose to normalize
the spectrum with respect to the pion decay
constant $F$. However, the renormalization of $F$ was only done
perturbatively, resulting in an estimated
uncertainty of about 8~\%. Here we determine
the renormalization constants non-perturbatively.
For instance, $Z_A$ turns out almost 10~\% smaller
than our previous estimate.
We also perform the continuum limit for $N=7$. Other values
of $N$ are in progress which will then enable a joint
large-$N$ and continuum limit extrapolation.

\section{Simulation details}
We employ the standard Wilson action for the gauge fields
and for the fermions.
In our main data set
we tune the lattice coupling, keeping the square root of the
string tension $a\sqrt{\sigma}\approx 0.2093$ 
fixed, in lattice units $a$.
In addition to this main infinite-$N$ extrapolation trajectory,
we now also
realize one finer lattice spacing
$a=0.1500/\sqrt{\sigma}$ and two coarser spacings
$a=0.2512/\sqrt{\sigma}$ and $a=0.3140/\sqrt{\sigma}$,
to enable a controlled continuum limit extrapolation.

The lattice 't~Hooft
coupling $\lambda=Ng^2=2N^2/\beta$
varies along the above trajectories of \emph{constant
physics}, i.e., constant lattice spacing in units
of the string tension, by terms of $\mathcal{O}(1/N^2)$. Other strategies,
e.g., keeping the pion decay constant in the chiral limit
$F$, the critical temperature $T_c$,
the gradient flow scale $t_0$ or $\lambda$ fixed,
are admissible, with specific advantages and disadvantages.
Fixing $\lambda$ for instance would be much more expensive
in terms of computer time as
two phase transitions have to be avoided along the
extrapolation to $N=\infty$: if $\lambda$ is taken
too large the system will undergo a strong coupling phase
transition once $a\sqrt{\sigma}\gtrsim 0.4$
while for small volumes $\ell^3\times 2\ell$,
i.e.\ for
$\ell<\ell_c\approx 2/\sqrt{\sigma}\gtrsim 1/T_c$,
a transition (similar to the finite temperature
transition) into a de-confined phase will occur.
We find $\lambda$ to reduce
by about 8~\% at constant $a\sqrt{\sigma}$,
when increasing $N$ from $N=3$ to $N=17$.
This means setting $\lambda$ sufficiently small to avoid
crossing into the strong coupling phase at large $N$ implies
tiny values of $a\sqrt{\sigma}\sim \exp\left[-1/(2b_0\lambda)\right]$
at small $N$, and hence of the lattice spacing $a$. This in turn
would necessitate a large number of
lattice points $N_s^3\times 2N_s\propto 1/a^4$
to remain in the confined phase $\ell=N_sa>\ell_c$.

We remark that, as long as $\ell>\ell_c$, finite volume
effects are irrelevant for the
large-$N$ extrapolation since these are suppressed
by factors $1/N^2$~\cite{Narayanan:2003fc}.
Nevertheless, due to the unitarity violations
of the quenched model, at small values of $N$ the volume
needs to be taken much bigger than this limit to enable
simulating light pion masses
down to $m_{\pi}\approx\sqrt{\sigma}/2$.

We cancel the leading $N$-dependence of meson decay constants
by defining
\begin{equation}
\hat{F}_{\pi}=
\sqrt{\frac{3}{N}}F_{\pi}\,,\quad
\hat{f}_{\rho}=
\sqrt{\frac{3}{N}}f_{\rho}\,.
\end{equation}
The normalization is chosen such that
$\hat{F}_X=F_X$ for $N=3$. $F_X=f_X/\sqrt{2}$ as usual.
We denote the (appropriately normalized) pion decay constant in the combined
chiral and large-$N$ limit as
\begin{equation}
\hat{F}_{\infty}=\lim_{N\rightarrow\infty}\hat{F}_{\pi}(m_q=0)
=\lim_{N\rightarrow\infty}\sqrt{\frac{3}{N}}F:=85.9(1.2)\,\mbox{MeV}\,,
\end{equation}
where we impose
the phenomenological QCD value~\cite{Colangelo:2010et}.
This gives a
lattice spacing $a\approx 0.095$~fm for $N\rightarrow\infty$,
along our main $a\sqrt{\sigma}=0.2093$ trajectory.
Of course we can only determine ratios of dimensionful quantities
and --- in the absence of experimental input from a
$N=\infty$ world --- any scale-setting in physical units
will be arbitrary and is just meant as a guide.
Nevertheless, we remark that
other ways of setting the scale appear to give similar results.
For instance, using the \emph{ad hoc} value $\sigma=1$~GeV/fm~$\approx
(444\,\mathrm{MeV})^2$, 
our lattice spacing reads $a\approx 0.093$~fm, instead.

We realize spatial extents $N_sa=24a\approx 2.3\,\mathrm{fm} \gg \ell_c$ at
$a\sqrt{\sigma}=0.2093$. To investigate finite
size effects, we also simulate $N_s=16$ and $N_s=32$ for
SU(2) and SU(3). For SU(3) no significant effects are found
and we conclude that our $N\geq 3$ results effectively
agree with the infinite volume limit.
At the two coarser spacings we simulate $N_s=16$ and $N_s=20$,
keeping the volume
approximately constant in physical units,
while we employ $N_s=24$ and $N_s=32$ on the finest lattice.
The largest $N=17$ we only simulate
at our main lattice spacing and restrict ourselves to
$N_s=12$. These small volume SU(17) results are found to
be consistent with the large-$N$ extrapolations of
the $7\geq N\geq 3$ data~\cite{Bali:2013fpo}, confirming
finite volume effects to become irrelevant at
large $N$ and also adding credibility to our extrapolation.
To enable chiral extra- and interpolations,
at each $N$ we realize at least six quark masses,
tuned to keep one set of pion masses approximately constant across the
different $\mathrm{SU}(N)$ theories and lattice spacings.
These correspond to pseudoscalar masses ranging from
$m_{\pi}\approx 2.7\sqrt{\sigma}$ down to
$m_{\pi}\approx 0.5\sqrt{\sigma}$ for $N\geq 5$ and
$m_{\pi}\approx 0.75\sqrt{\sigma}$ for $N\leq 4$.

\FIGURE[t]
{
\includegraphics[width=.47\textwidth,clip]{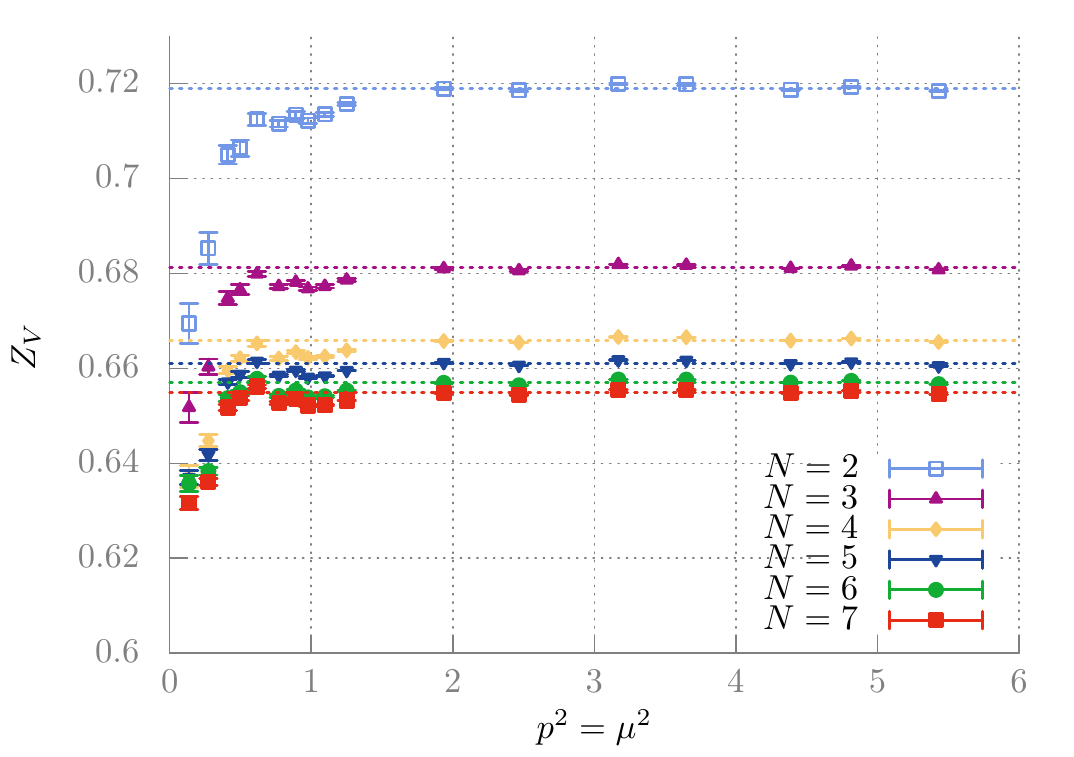}~~
\includegraphics[width=.47\textwidth,clip]{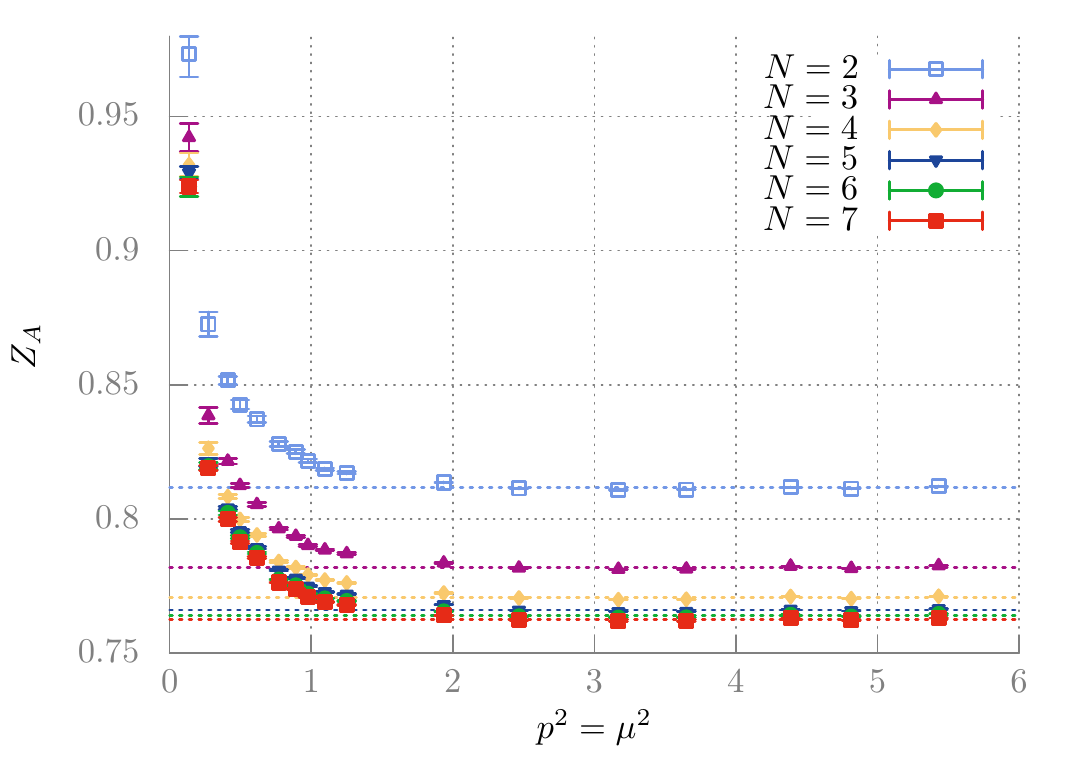}
\caption{Non-perturbative determination of the
vector and axial-vector renormalization constants.\label{fig:renorm1}}}

\section{Renormalization constants}

\FIGURE[ht]
{
\includegraphics[width=.6\textwidth,clip]{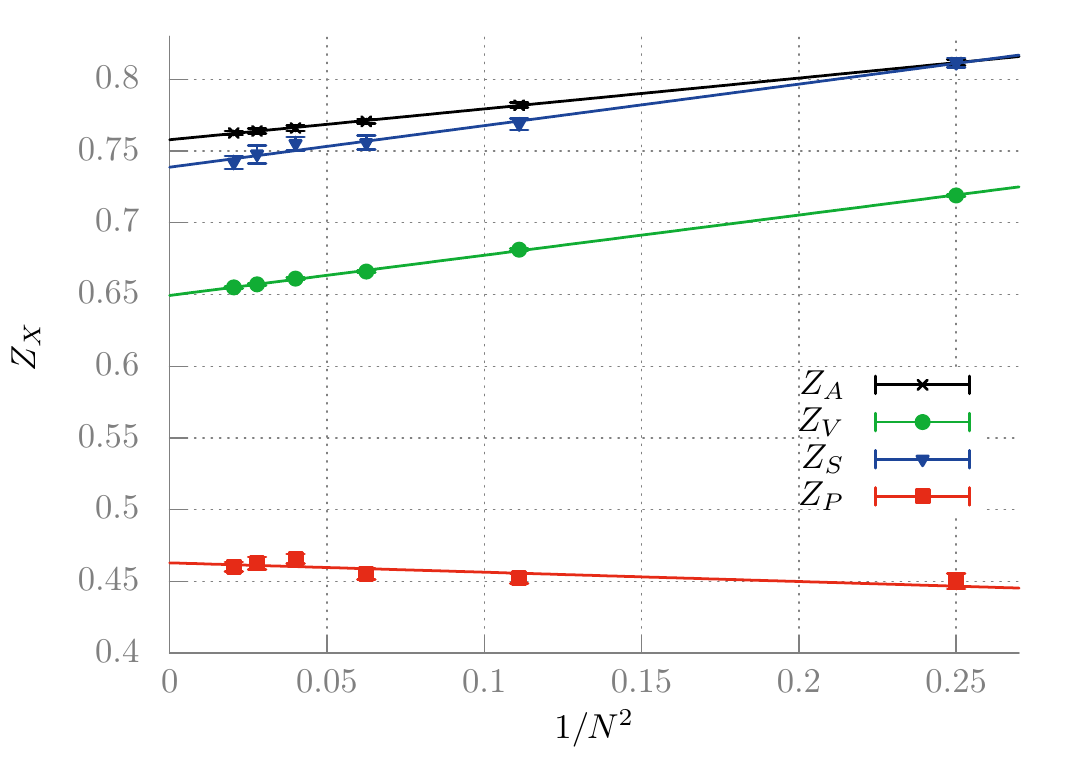}
\caption{Renormalization factors versus $1/N^2$. The
scalar and pseudoscalar factors $Z_S$ and $Z_P$ translate the lattice
results obtained at $a^{-1}\approx 2.13$~GeV
to the $\overline{\mathrm{MS}}$ scheme at a scale
$\mu = 2$~GeV~$\approx 23.3\,\hat{F}_{\infty}$.\label{fig:renorm2}}}

The hopping parameter $\kappa$ is related to the
vector and axial quark Ward identity lattice quark masses $m_q$
and $m_{{}_\mathrm{PCAC}}$, respectively, via
\begin{equation}
\label{eq:bare_mass}
am_q=\frac{1}{2}\left(\frac{1}{\kappa}-\frac{1}{\kappa_c}\right)=
\frac{Z_AZ_S}{Z_P}\left(1-b\,am\right)am_{{}_\mathrm{PCAC}}\,,
\end{equation}
where $\kappa=\kappa_c$ corresponds to a massless quark.
The improvement parameter $b$ is, for our calculation with
unimproved Wilson quarks, redundant. Fitting
$am_{{}_\mathrm{PCAC}}$  for each $N$
as a function of $1/\kappa$ according to the above
parametrization, we obtain
the critical hopping parameters $\kappa_c(N)$ and the
(scale-independent) combination of renormalization
factors $Z_AZ_S/Z_P$, as described in ref.~\cite{Bali:2013fpo}.
We use this to determine
$Z_P(a\mu)$, once $Z_A(a)$ and $Z_S(a\mu)$ have been computed.

We determine the renormalization constants
$Z_A(a)$ (required for the pion decay constant) and
$Z_V(a)$ (for the vector decay constant) via the
Roma-Southampton non-perturbative
matching~\cite{Martinelli:1994ty} to the RI'MOM scheme.
In the case of $Z_S(a\mu)$ (needed for the chiral
condensate and quark mass renormalization, not presented here)
this is then perturbatively matched to the $\overline{\mathrm{MS}}$ scheme.
To remove lattice artefacts we
parameterize (see, e.g., ref.~\cite{Constantinou:2013ada}):
\begin{equation}
Z_X(a)=Z_X^{\mathrm{latt}}(p,a)-z_0(a)S^2(ap)-z_1(a)\frac{S^4(ap)}{S^2(ap)}\,,
\end{equation}
where $S^n(ap):=\sum_{\mu}(ap_{\mu})^n$. $Z_X$, $z_0$ and $z_1$ are fit
parameters. In the case of $Z_S$ which has an anomalous dimension,
we take $S^2(ap)$ as the argument of the leading log as indicated
by lattice 
perturbation theory.
The lattice artefact subtracted data for $Z_V$ and $Z_A$ are displayed
for the various $N$-values in figure~\ref{fig:renorm1}
and the four renormalization factors are shown in
figure~\ref{fig:renorm2}.

\section{Spectrum and decay constants}

We compute correlation matrices between differently
smeared interpolators. This gives us access to excited states in
many channels, in addition to the ground states.
We then perform joint large-$N$ and chiral extrapolations.
As demonstrated in ref.~\cite{Bali:2013fpo}, the
$N\geq 3$ data are consistent with purely quadratic
dependencies on $1/N$, with small slopes. A notable exception
is the scalar particle $a_0$.
The chiral extrapolations are performed as
polynomials in the quark mass $m_{{}_\mathrm{PCAC}}$ that can be
determined more precisely than $m_{\pi}^2$.
For $N\leq 5$, we detect the expected chiral log.

\FIGURE[ht]
{
\includegraphics[width=.6\textwidth,clip]{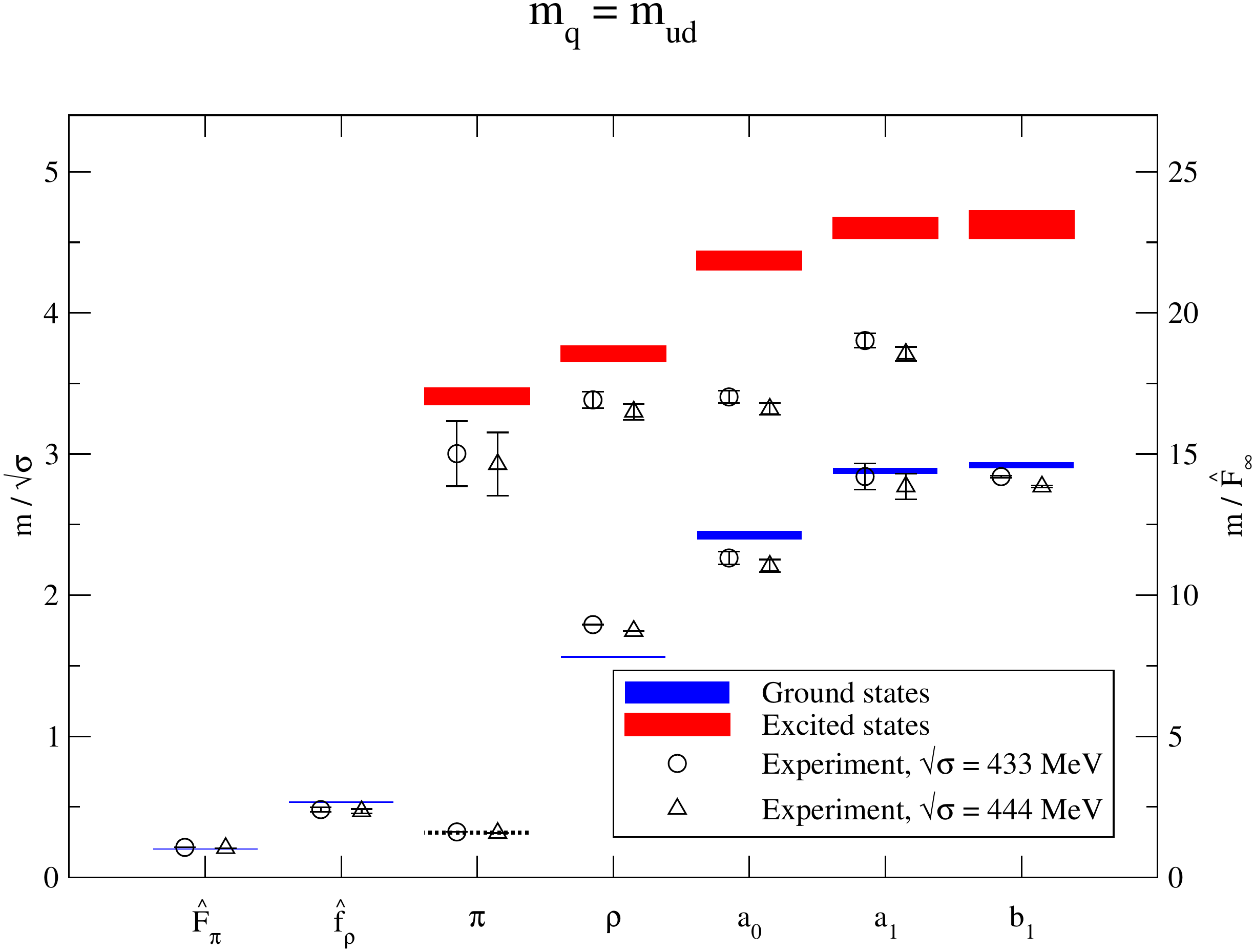}
\caption{The SU($\infty$) spectrum at
$a=0.2093/\sqrt{\sigma}\approx 0.095$~fm
at the physical light quark mass ($m_{\pi}\approx 1.6\,\hat{F}_{\infty}$).
\label{fig:spectrum}}}

We interpolate and extrapolate the spectrum to three values
of the quark mass, $m=0$, $m=m_{ud}$ and $m=m_s$, where
$m_{\pi}(m_{ud})=138\,\mathrm{MeV}\approx 1.6\, \hat{F}_{\infty}$
and $m_{\pi}(m_s)=\left(m_{K^{\pm}}^2+m^2_{K^0}-m_{\pi^{\pm}}^2\right)^{1/2}=
686.9\,\mathrm{MeV}\approx 8.0\, \hat{F}_{\infty}$.
We display the resulting $m=m_{ud}$ spectrum
for the lattice spacing $a\approx 0.095$~fm in figure~\ref{fig:spectrum}.
On the scale of the plot this is indistinguishable from the
$m=0$ spectrum. The value $\sqrt{\sigma}=433$~MeV corresponds to
setting the scale with $\hat{F}_{\infty}$ as the input.
Due to the non-perturbative renormalization of the
pion decay constant, this differs somewhat from our previous
results~\cite{Bali:2013fpo}.
Interestingly, the ground states, including the $a_0$, are close to the
experimental $N=3$ QCD values. However, the continuum limit still needs
to be taken.

\FIGURE[t]
{
\includegraphics[width=.7\textwidth,clip]{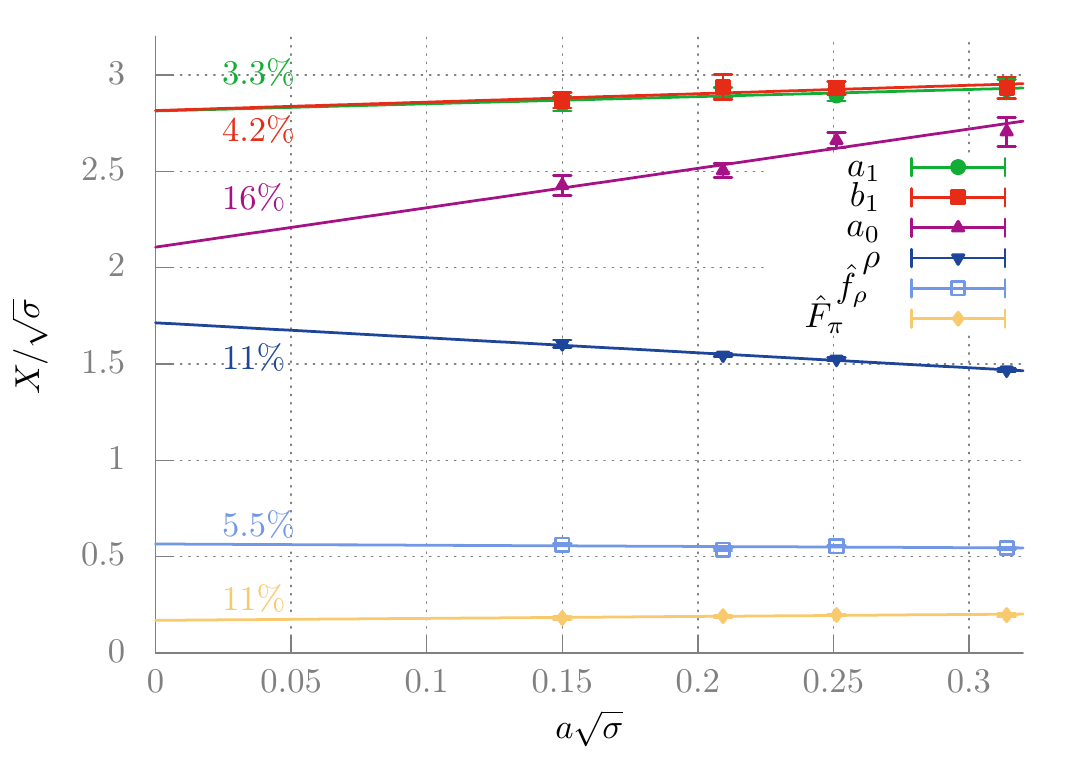}
\caption{Continuum limit extrapolation of the SU(7) results.
The percentage numbers indicate changes relative to
the results obtained at $a\sqrt{\sigma}=0.2093$
(second data points from the left).\label{fig:continuum}}}

We are in the process of performing a combined large-$N$
and continuum limit. Within errors the finite-$a$ SU(7) results
agree with our $N\rightarrow\infty$ extrapolations.
In figure~\ref{fig:continuum}
we display the continuum limit extrapolation of some SU(7)
masses and decay constants.
Quantitatively, the slopes are very similar to
results obtained previously in the SU(3)
theory~\cite{Butler:1994em,Aoki:2002fd}.
Therefore, we do not anticipate complications
when the combined limit will be performed. It is clear
from the extrapolation that the finite-$a$ masses displayed in
figure~\ref{fig:spectrum} are subject to systematics of
order 10~\%. In particular, 
the ratio $m_{a_0}/m_{\rho}$ will move closer to unity than
that figure suggests.

\section{Summary}
We have determined the decay constants as well as
the ground and first excited
state masses of mesons in the large-$N$ limit of QCD. A continuum
limit extrapolation is in progress. This
will then allow the results to be used as input, e.g., to effective
field theory calculations.

\acknowledgments
\begin{sloppypar}
\noindent
This work is supported by the EU ITN STRONGnet (grant 238353),    
by the German DFG (SFB/TRR 55), by the UK STFC (grant
ST/G000506/1), by the Royal Society (grant UF09003)
and by the Spanish MINECO (grant SEV-2012-0249).
The simulations were performed on the Regensburg
iDataCool cluster, at LRZ Munich, on High Performance Computing Wales
systems and at the Finnish IT Center for Science (CSC), Espoo.
The code is based on the Chroma suite~\cite{Edwards:2004sx}.
\end{sloppypar}

\end{document}